\documentclass[12pt]{article}
\usepackage{graphicx}
\renewcommand{\baselinestretch}{1.}
\begin{document}
\begin{center}
{\Large\bf A study of transverse momentum distributions of jets
produced in $p$-$p$, $p$-$\bar p$, $d$-Au, Au-Au, and Pb-Pb
collisions at high energies}

\vskip1.0cm

Hua-Rong Wei, Fu-Hu Liu{\footnote{E-mail: fuhuliu@163.com;
fuhuliu@sxu.edu.cn}}

{\small\it Institute of Theoretical Physics, Shanxi University,
Taiyuan, Shanxi 030006, China}

\end{center}

\vskip1.0cm

{\bf Abstract:} The transverse momentum distributions of jets
produced in $p$-$p$, $p$-$\bar p$, $d$-Au, Au-Au, and Pb-Pb
collisions at high energies with different selected conditions are
analyzed by using a multi-source thermal model. The
multi-component (mostly two-component) Erlang distribution used in
our description is in good agreement with the experimental data
measured by the STAR, D0, CDF II, ALICE, ATLAS, and CMS
Collaborations. Related parameters are extracted from the
transverse momentum distributions and some information on
different interacting systems are obtained. In the two-component
Erlang distribution, the first component has usually two or more
sources which are contributed by strong scattering interactions
between two quarks or more quarks and gluons, while the second
component has mostly two sources which are contributed by harder
head-on scattering between two quarks.
\\

{\bf Keywords:} High-energy jets, Transverse momentum
distribution, Erlang distribution, Hard scattering, Effective
temperature
\\

PACS Nos.: 25.75.Ag, 24.10.Pa, 25.75.Dw
\\

{\section{Introduction}}

As a new state of matter, quark-gluon plasma (QGP) [1] is
different from common plasmas. Generally, the QGP can not be
observed directly because its creation needs an extremely high
temperature and density. In order to study the properties of the
QGP and the mechanisms of parton interactions, a lot of
experiments of high energy heavy ion collisions have been
performed. Particularly, a high density and high temperature
location is formed at the Relativistic Heavy Ion Collider (RHIC)
and the Large Hadron Collider (LHC) to create the QGP and to
produce multiple final-state particles. By investigating and
analyzing these particles, one can obtain some information on the
QGP. Hadron jets are one type of final-state particles. People
study jets production process to learn the properties of the QGP
and the mechanisms of parton interactions both in theories and in
experiments [2--4].

The transverse momenta of particles carry important information
about the dynamics of particle productions and the evolution
process of interacting system formed in nucleus-nucleus collisions
[5]. The transverse momentum ($p_T$) distributions of final-state
particles reflect the state of the interacting system at the
kinetic stage of freeze-out, when hadrons are no longer
interaction and their momenta do not change [6]. Comparing to the
low-$p_T$ hadrons which are originated from strong interactions
among gluons and sea quarks, high-$p_T$ hadrons are produced due
to harder head-on scattering between valent quarks. To analyze
high-$p_T$ spectra of final-state particles is then important in
studying the mechanisms of high-energy collisions. In addition,
the mean transverse momentum of particles, served as a probe for
the equation of state of the hadronic matter [7], can partly
reflect the effective temperature of interacting system and the
transverse excitation degree of emission source in high energy
collisions [8].

Due to the hard-scattering of incident partons, partonic jets
(including quark and gluon jets) materialize very early during the
collisions, which result in plenty of outgoing high-$p_T$ partons.
Soon afterwards, a parton fragments into a leading hadron, and a
large number of which gather into a cone of hadronic jets which
are detected by various detectors such as STAR, D0, CDF II, ALICE,
ATLAS, and CMS [9--36]. We are interested in analyzing the
transverse momentum distributions of various jets because they can
reflect some information of early collisions of partons.

In this paper, we use the multi-component (mostly two-component)
Erlang distribution to describe the transverse momentum
distributions of jets produced in $p$-$p$, $p$-$\bar{p}$, $d$-Au,
Au-Au, and Pb-Pb collisions at high energies, with the framework
of a multi-source thermal model [37--39]. The quoted data are at
five energies: 0.2, 1.96, 2.76, 7, and 8 TeV per nucleon pair
[9--36], and were selected in different detection channels
according to different conditions of cone radius, integrated
luminosity, pseudorapidity (rapidity) range, and centrality
interval. The calculations are performed by using the Monte Carlo
method.
\\

{\section{The model and formulism}}

According to the multi-source thermal model [37--39], we assume
that many emission sources are formed in high energy collisions.
These emission sources can be separated into a few groups due to
different interacting mechanisms in the collisions and different
event samples in experimental measurements. Each emission source
in the same group has the same transverse excitation degree, and
the group is assumed to stay at a local equilibrium state. The
total result of a multi-group emission process contributes to the
final distribution.

The transverse momentum distribution generated in thermodynamic
system obeys an exponential distribution, which is the
contribution of one emission source. We have
\begin{equation}
f_{ij}(p_{tij})=\frac{1}{\langle{p_{tij}}\rangle}\exp{\bigg[
-\frac{p_{tij}}{\langle{p_{tij}}\rangle}\bigg]},
\end{equation}
where $p_{tij}$ is the transverse momentum contributed by the
$i$-th source in the $j$-th group, and $\langle{p_{tij}}\rangle$
is the mean value of different $p_{tij}$. The folding result of
$m_j$ sources in a given (the $j$-th) group is an Erlang
distribution, i.e.
\begin{equation}
f_{j}(p_{T})=\frac{p_T^{m_{j}-1}}{(m_{j}-1)!\langle{p_{tij}}\rangle^{m_{j}}}
\exp{\bigg[-\frac{p_{T}}{\langle{p_{tij}}\rangle}\bigg]},
\end{equation}
where $p_{T}$ is the transverse momentum contributed by the
$m_{j}$ sources. If the number of groups is $l$, the total
contribution is the multi-component Erlang distribution expressed
as
\begin{equation}
f(p_{T})=\sum_{j=1}^{l}k_{j}f_{j}(p_{T}),
\end{equation}
where $k_{j}$ is the weight contributed by the $j$-th group, which
obeys the normalization $\sum k_j=1$. To describe transverse
excitation degree of emission source in the present work, we
define a new effective temperature parameter
\begin{equation}
T\equiv\sum_{j=1}^{l}k_{j}\langle{p_{tij}}\rangle,
\end{equation}
which uses the inverse slope parameter $\langle{p_{tij}}\rangle$.

We shall see from the following section that $p_T$ distributions
of jets analyzed in the present work can be mostly fitted by the
two-component Erlang distribution. Generally, the first component
corresponds to strong scattering interactions where two quarks or
more quarks and gluons take part in the process, and the second
one corresponds to harder head-on scattering where two quarks take
part in the process. All the calculations on $p_T$ distributions
are performed by using the Monte Carlo method. For the first
component, we have
\begin{equation}
p_T=-\langle p_{ti1} \rangle \sum_{i=1}^{m_1} \ln R_i,
\end{equation}
and for the second component, we have
\begin{equation}
p_T=-\langle p_{ti2} \rangle \sum_{i=1}^{m_2} \ln R_i,
\end{equation}
where $R_i$ denote random numbers in $[0,1]$. The statistics based
on weights $k_1$ and $k_2$ ($=1-k_1$) will give $p_T$
distributions. In the case of using the three- or multi-component
Erlang distribution, the calculation process of the Monte Carlo
method is similar.

We would like to point out that the new effective temperature
parameter defined in Eq. (4) is not the real temperature of jets
production source, but the mean transverse momentum extracted from
the model or experimental data. The real temperature of jets
production source should be a reflection of purely thermal motion
in the source. The effect of flow or blast wave should be excluded
in the extraction of source temperature. Generally, the effective
temperature parameter is extracted from the transverse momentum
distribution. It includes thermal motion and flow effect, and
should be greater than the source temperature. As the mean
transverse momentum, the new effective temperature parameter is
independent of models.
\\

{\section{Comparison with experimental data}}

Figure 1 presents $p_T$ distributions of high tower (HT) trigger
jets in $p$-$p$ collision (a), HT trigger jets in 0--20\% central
Au-Au collisions (b), jets from $d$-Au collisions (c), and
uncorrected jets in $p$-$p$ collision (d) at center-of-mass energy
$\sqrt{s_{NN}}=0.2$ TeV, where for $p$-$p$ collision
$\sqrt{s_{NN}}$ can be simplified as $\sqrt{s}$. The symbols
represent the experimental data of the STAR Collaboration [9--11].
In Figs. 1(a) and 1(b), jet events were selected by an online HT
trigger using the anti-$k_{T}$ algorithm [12] with a cone radius
$R=0.4$, pseudorapidity range $|\eta|<0.6$, and $p_T>2.0$ GeV/$c$.
The displayed uncertainty of the data is only systematic
uncertainty. The data from Fig. 1(c) were recorded during RHIC run
8 (2007--2008) with $R = 0.4$ and $|\eta|<0.55$. The error bars
represent systematic uncertainty and statistical uncertainty. Raw
jets in Fig. 1(d) were obtained under the condition of $R=0.7$,
$-0.7<\eta<0.9$, and jets $p_T>5.0$ GeV/$c$. In the figure, some
error bars are not visible due to their small values. In this case
we regard the marker size as the errors in the calculation. The
curves are our results calculated with the two-component Erlang
distribution in the framework of the multi-source thermal model by
using the Monte Carlo method. The values of free parameters,
defined effective temperature parameter $T$, and $\chi^2$ per
degree of freedom ($\chi^2$/dof) or $\chi^2$ are shown in Table 1,
where $\chi^2$ is used in the case of the number of data points
being less than 6 which is the number of free parameters and
normalization constant. One can see that the multi-source thermal
model describes the experimental data of the considered jets in
$p$-$p$, $d$-Au, and Au-Au collisions at $\sqrt{s_{NN}}=0.2$ TeV.
The transverse momentum distributions of mentioned jets are shown
to obey the two-component Erlang distribution. For Figs. 1(a),
1(b), and 1(c), all the source numbers of the second groups are 2
and the values of mean excitation degree (defined effective
temperature parameter) $T$ extracted from the spectra are close to
each other. For the raw jets in Fig. 1(d), the background
interference leads to the number of emission sources to increase
and the mean excitation degree to decrease.

Figure 2 shows $p_T$ spectra of the leading jets (a)--(c) and all
jets (d) that satisfy the event selection described below,
produced in $p$-$\bar{p}$ collision at $\sqrt{s}=1.96$ TeV. The
symbols in (a) and (b) represent the experimental data collected
by the D0 Collaboration in Run II with an integrated luminosity of
9.7 fb$^{-1}$ [13], while (c) for 3.7 fb$^{-1}$ [14], and (d) the
CDF II Collaboration with a cone size of $R=0.7$, $0.1
<|\eta|<0.7$, as well as an integrated luminosity of 6 fb$^{-1}$
[15]. The data come from the $Z$ bosons decaying to pairs of
leptons $ee$ (a) or $\mu\mu$ (b). The jets in Fig. 2(c) are
required to satisfy $p_T > 20$ GeV/$c$ and rapidity $|y|<3.2$ with
a cone radius $R = 0.5$. The error bars in (a) and (b) represent
the sum of statistical and systematic uncertainties, and those in
(c) and (d) represent statistical and systematic uncertainties
respectively. In the figure, for the points with invisible error
bars, we measure the marker size as the errors used in the
calculation. The curves are our results calculated with the
two-component Erlang distribution. The values of free parameters,
defined $T$, and $\chi^2$/dof in the calculation are given in
Table 1. One can see that the two-component Erlang distribution
describes well the experimental data of the jets produced in
$p$-$\bar{p}$ collision at $\sqrt{s}=1.96$ TeV. For the four
cases, all the source numbers of the second groups are 2, and the
values of $T$ extracted from the spectra are approximately the
same.

In Figure 3, we give charged jet $p_T$ spectra produced in Pb-Pb
collisions at $\sqrt{s_{NN}}=2.76$ TeV in different centrality
intervals of 0--10\% (a), 10--30\% (b), 30--50\% (c), and 50--80\%
(d). The experimental data represented by the symbols were
measured with a cone radius $R=0.2$, $|\eta|<0.5$, anti-$k_T$ jet
algorithm [12], charged track $p_T>0.15$ GeV/$c$, and leading
charged track $p_T>10$ GeV/$c$ by the ALICE Collaboration [16].
The error bars include the total uncertainty. The curves are our
fitted results with the two-component Erlang distribution. The
values of free parameters, defined $T$, and $\chi^2$/dof in the
calculation are displayed in Table 1. Once more, the two-component
Erlang distribution describes well the experimental data of the
charged jets produced in Pb-Pb collisions at $\sqrt{s_{NN}}=2.76$
TeV with different centrality intervals. The fitted source numbers
of the second groups are also 2, and the effective temperature
parameter $T$ extracted from the spectra slightly increases with
increase of the centrality percentage $C$, or $T$ has no obvious
change with increase of $C$ in the error range. The figure display
of the dependence of $T$ on $C$ will be discussed in the last part
of this section.

Figure 4 displays jet $p_T$ spectra produced in $p$-$p$ collision
at $\sqrt{s}=7$ TeV for different conditions. The symbols in (a)
and (b) represent the experimental data recorded by the CMS
Collaboration [17], and those in (c) and (d) represent the ATLAS
Collaboration [18--19]. The first [the black squares in Fig. 4(a)]
and the second leading jets [the red circles in Fig. 4(a)], as
well as the lepton+jets channel jets with $p_T>35$ GeV/$c$ [the
black squares in Fig. 4(b)] and the dilepton channel jets with
$p_T>30$ GeV/$c$ [the red circles in Fig. 4(b)], were selected
corresponding to a total integrated luminosity of 5.0 fb$^{-1}$.
In Fig. 4(c), the selected data for the Cambridge-Aachen algorithm
jet events where the number of reconstructed primary vertices
($N_{PV}$) composed of at least five tracks is exactly one
corresponds to an integrated luminosity of 2 pb$^{-1}$, $R=1.2$,
and $|y|<2$. The jets shown by the red circles in Fig. 4(c) were
split and filtered from those jets with $p_T>200$ GeV/$c$
represented by the black squares in the same panel. The data for
$b$-quark jets ($b$-jets) displayed in Fig. 4(d) were collected
with an integrated luminosity of 1.8 fb$^{-1}$ for the
single-lepton channels by the black squares and for the dilepton
channels by the red circles. The error bars on the data points
indicate the statistical uncertainty in Figs. 4(a)--4(c) and
systematic uncertainty in Fig. 4(d). For some data points in which
the error bars are invisible in the figure, we use the marker size
instead of the real error values in our calculation. The curves
are our fitted results by the two-component Erlang distribution.
The relevant parameter values with $\chi^2$/dof in the calculation
are presented in Table 1. The two-component Erlang distribution
with $m_{2}=2$ provides a good description on the data. As can be
seen, the value of $T$ extracted from the $1^{st}$ leading jet
spectra is significantly larger than that extracted from the
$2^{nd}$ one. In the error range, $T$ extracted from the
single-lepton channel jet spectra is approximately in agreement
with that from the dilepton one.

The transverse momentum distributions for the subleading $b$-jets
after the $Z+2b$-jets selection corresponding to an integrated
luminosity of 5.0 fb$^{-1}$, the soft-muon tagging (SMT) jets for
opposite-sign and same-sign (OS-SS) events in $W$+1,2 jets samples
in electron channels with an integrated luminosity of 4.6
fb$^{-1}$, and the light-quark jets produced in the same condition
as that in Fig. 4(d) in $p$-$p$ collision at $\sqrt{s}=7$ TeV are
shown in Figs. 5(a)--5(c) respectively. The experimental data
represented by the symbols were recorded by the CMS Collaboration
[20] and the ATLAS Collaboration [19, 21]. The error bars indicate
the total uncertainty. In Fig. 5(c), the error bars of the points
in high-$p_T$ area present the marker size of the points from Ref.
19, in which the linear coordinate is used. The curves are our
results calculated by the two-component Erlang distribution. The
parameter values and the defined $T$ in the calculation are shown
in Table 1 with values of $\chi^2$/dof. One can see that the
experimental data are in good agreement with the two-component
Erlang distribution with $m_{2}=2$. In the error range, the three
effective temperatures extracted from the three distributions are
approximately equal to each other.

Figure 6 shows some other jet $p_T$ distributions in $p$-$p$
collision at $\sqrt{s}=7$ TeV. The symbols in (a)--(d) represent
the experimental data which were recorded with the ATLAS detector
corresponding to an integrated luminosity of 37 pb$^{-1}$ [22],
the CMS detector corresponding to an integrated luminosity of 5.0
fb$^{-1}$ for the leading $b$-tagged jets in the $Z+2b$-jets
sample [20], the ATLAS detector corresponding to an integrated
luminosity of 4.6 fb$^{-1}$ for the leading $b$-tagged jets in
$e$+jets channels [23], and the ATLAS detector corresponding to an
integrated luminosity of 4.6 fb$^{-1}$ for the leading $b$-tagged
jets in $\mu$+jets channels [23], respectively. The statistical or
systematic uncertainties are included in the error bars. For some
data points with no error bars, we use the marker size as their
errors in the calculation. The Monte Carlo results calculated by
using the two-component Erlang distribution are indicated by the
curves, and the corresponding parameters, defined $T$, and
$\chi^2$/dof are listed in Table 1. The fitting distributions are
in agreement with the experimental data. All the source numbers of
the second components are 2. The values of $T$ extracted from the
leading jets in Figs. 6(b)--6(d) are larger than that extracted
from inclusive jets in Fig. 6(a). In addition, for the leading
$b$-tagged jets, $T$ from $e$+jets channels is in agreement with
that from $\mu$+jets channels within uncertainties.

The reconstructed jet $p_T$ spectra for the leading (squares),
$2^{nd}$ (circles), $3^{rd}$ (triangles), $4^{th}$ (stars), and
$5^{th}$ order jets (diamonds) in the electron ($e$+jets) channels
produced in $p$-$p$ collision at $\sqrt{s}=7$ TeV are indicated in
Fig. 7. The symbols represent the experimental data collected by
the ATLAS Collaboration corresponding to an integrated luminosity
of 4.6 fb$^{-1}$ and a radius parameter of $0.4$ within the
pseudorapidity range $|\eta| < 2.5$ using anti-$k_T$ jet algorithm
[12] for $p_T>25$ GeV/$c$ jets [24]. In the figure, all the data
have invisible error bars (statistical uncertainty only), so we
measure the marker size as the errors used in the calculation. The
curves are our fitted results with the two-component Erlang
distribution, and the related parameter values with $\chi^2$ are
given in Table 1. From the figure one can see that the fitted
results are in agreement with the observed jet $p_T$
distributions. For different jet orders, we have $m_1=7$ and
$m_2=2$. The value of $T$ decreases with increase of the jet order
$O$. The figure display of the dependence of $T$ on $O$ will be
discussed in the last part of this section.

In Figure 8, the leading jet $p_T$ distributions in (a) the 4-jet
$t \bar{t}$ , (b) preselection, (c) $W$ boson+jets, and (d) 5-jet
$t\bar{t}$ with an integrated luminosity of 4.7 fb$^{-1}$ in
$p$-$p$ collision at $\sqrt{s}=7$ TeV are shown. The symbols
represent the experimental data recorded by the ATLAS
Collaboration [25]. The error bars indicate the total uncertainty.
The curves indicate the fitted results by the two-component Erlang
distribution, and the corresponding values of free parameters,
defined $T$, and $\chi^2$/dof are listed in Table 1. One can see
that the two-component Erlang distribution with $m_2=2$ describes
the data of the leading jet $p_T$ spectra, and the extracted
effective temperature parameters have a smaller difference.

Figures 9(a) and 9(b) display $p_T$ distributions of the leading
and fat jets in $p$-$p$ collision at $\sqrt{s}=7$ TeV
respectively, where the results are in the Lepton Plus Jets
Channels using an integrated luminosity of 2.05 fb$^{-1}$. Figures
9(c)--9(f) give $p_T$ distributions of the leading jets in the
same collision, and the corresponding integrated luminosity and
lepton plus jets channels are shown in the panels. The symbols
represent the experimental data of the ATLAS Collaboration
[26--30]. The error bars indicate the experimental uncertainty.
Our calculated results by using the two-component Erlang
distribution are presented by the curves in the figure. The values
of free parameters, defined $T$, and $\chi^2$/dof are given in
Table 1. One can see that the mentioned $p_T$ spectra for the
leading and fat jets are fitted by the two-component Erlang
distribution with $m_2=2$ and relatively larger $T$. In addition,
in the error range, $T$ extracted from the jet spectra in $e$+jets
channels is in agreement with that in $\mu$+jets channels.

The transverse momentum distributions of the leading jets produced
in $p$-$p$ collision at $\sqrt{s}=8$ TeV with an integrated
luminosity of 14.3 fb$^{-1}$ in the $e$+jets and an integrated
luminosity of 14.2 fb$^{-1}$ in the $\mu$+jets, after the resolved
selection, are shown in Figs. 10(a) and 10(b) respectively. The
experimental data (the squares) were recorded by the ATLAS
Collaboration [31]. Meanwhile, the data (the squares) of the
leading light jet $p_T$ distribution and the second leading light
jet $p_T$ distribution from $p$-$p$ collision at $\sqrt{s}=8$ TeV
corresponding to a total integrated luminosity of 19.5 fb$^{-1}$
measured by the CMS Collaboration [32] are presented in Figs.
10(c) and 10(d) respectively. Moreover, the leading jet $p_T$
distribution and the subleading jet $p_T$ distribution in the
dijet system using $p$-$p$ collision at $\sqrt{s}=8$ TeV with an
integrated luminosity of 20.3 fb$^{-1}$ are displayed in Figs.
10(e) and 10(f) respectively, where the squares represent the data
of the ATLAS Collaboration [33]. In the figure, the error bars
reflect the statistical uncertainty, and the curves are our
results calculated by using the two-component Erlang distribution.
The relevant parameter values with $\chi^2$/dof in our calculation
are presented in Table 1. From the figure, one can see that the
modelling results are in agreement with the data. Comparing with
the leading jet spectra, the defined $T$ extracted from the
subleading jet spectra exhibits a quickly decrease. The defined
$T$ extracted from the light jet spectra shows relatively a little
larger than that from the non-tagged jets spectra, or the two
situations are approximately the same in the error range.
Moreover, the defined $T$ corresponding to the leading jets from
$e$+jets and $\mu$+jets channels in Figs. 10(a) and 10(b) are
nearly the same.

The inclusive jet $p_T$ distributions and the large-$R$ jet $p_T$
distribution produced in $p$-$p$ collision at $\sqrt{s}=8$ TeV are
presented in Figure 11. The data (squares) in Fig. 11(a) were
collected with the CMS detector corresponding to an integrated
luminosity of 10.7 fb$^{-1}$ from a high-level trigger that
accepted events containing at least one jet with $p_{T}>320$
GeV/$c$ [34]. The data (squares) in Fig. 11(b) for $Z \rightarrow
\mu\mu$ events were selected from the full 2012 run and amounted
to a total integrated luminosity of 19.8 fb$^{-1}$ [35]. The data
(squares) in Fig. 11(c) were collected by the ATLAS Collaboration
corresponding to an integrated luminosity of 20 fb$^{-1}$ with an
uncertainty of 2.8\% [36]. In the figure, the curves indicate our
fitted results. One can see that the modelling results are in good
agreement with the experimental data. The related parameter values
for Figs. 11(a) and 11(b) are given in Table 1, where the number
of components is two and the source numbers of the second
components are $m_{2}=2$. For Fig. 11(c), we have to use the
three-component Erlang distribution with $m_1=29$,
$p_{ti1}=(6.65\pm0.60)$ GeV/$c$, $k_1=0.396\pm0.036$, $m_2=14$,
$p_{ti2}=(19.50\pm1.00)$ GeV/$c$, $k_2=0.204\pm0.036$, $m_3=2$,
$p_{ti3}=(111.00\pm10.00)$ GeV/$c$, $T=51.011\pm11.853$ GeV, and
$\chi^2$/dof=0.377.

To see clearly the dependence of the effective temperature
parameter $T$ on the centrality percentage $C$ in Pb-Pb collisions
at $\sqrt{s_{NN}}=2.76$ TeV, we present the relation $T-C$ in Fig.
12(a). The circles represent the effective temperature values
obtained from Figure 3 and listed in Table 1, and the curve is our
fitted result by an exponential function of
\begin{equation}
T=(-0.125\pm0.020)\exp \bigg[- \frac{C}{(10.500\pm3.000)} \bigg]
+(5.878\pm0.010)
\end{equation}
with $\chi^2$/dof=0.0002, where $T$ is in the units of GeV and $C$
is in \%. One can see that $T$ slightly increases with increase of
$C$, and a saturation effect quickly appears in semi-central
collisions, or $T$ has no obvious change in the error range with
increase of $C$. In addition, to see clearly the dependence of the
effective temperature parameter $T$ on the jet order $O$ in
$p$-$p$ collision at $\sqrt{s}=7$ TeV, we present the relation
$T-O$ in Fig. 12(b). The circles represent the effective
temperature values obtained from Figure 7 and listed in Table 1,
and the curve is our fitted result by an exponential function of
\begin{equation}
T=(54.30\pm3.00)\exp \bigg[- \frac{O}{(1.43\pm0.10)} \bigg]
+(4.55\pm1.00)
\end{equation}
with $\chi^2$=0.129, where $T$ is in the units of GeV. One can see
that $T$ decreases with increase of $O$.

The dependences of the effective temperature parameter $T$ on
other factors such as $l$ and di-$l$ channels (a), $\mu(\mu\mu)$
and $e(ee)$ channels (b), leading and sub-leading jets (c), as
well as size of interacting system (d) are displayed in Figure 13,
where the symbols represent the effective temperature values
obtained from the above figures and listed in Table 1. One can see
that a high center-of-mass energy results in a high effective
temperature, different lepton channels show approximately the same
effective temperature in the error range, and the leading jets
correspond to a high effective temperature comparing with the
sub-leading jets. At the same time, $T$ slightly decreases with
increase of the size of interacting system, or $T$ is
approximately not related to the size in the error range. The
former case can be explained by the influence of jet quenching
effect (quickly energy loss) which exists in Au-Au collisions and
doesn't exist in $p$-$p$ collision.
\\

{\section{Conclusions}}

From the above discussions, we obtain following conclusions.

(a) The transverse momentum distributions, of various jets
produced in $p$-$p$, $p$-$\bar{p}$, $d$-Au, Au-Au, and Pb-Pb
collisions over an energy range from 0.2 to 8 TeV in different
additional selection conditions, are described by using the
multi-component Erlang distribution in the framework of the
multi-source thermal model which reflects the reaction types in
interacting system and multiple temperatures emission. The
calculated results are in good agreement with the experimental
data measured by the STAR, D0, CDF II, ALICE, ATLAS, and CMS
Collaborations.

(b) Except for one group data in Fig. 11(c) which is described by
the three-component Erlang distribution, the data of $p_T$ spectra
for mentioned jets are fitted by the two-component Erlang
distribution. The source numbers of the first components are
greater than or equal to 2, while the source numbers of the second
components are mostly 2. In $p_T$ distribution, the
first-component in charge of low-$p_T$ jets indicates strong
scattering interactions between two quarks or among more quarks
and gluons and accounted for a larger proportion, and the
second-component for high-$p_T$ region means harder head-on
scattering between two quarks.

(c) The effective temperature parameter $T$ is higher, which means
that the transverse excitation of emission source is very violent.
From our study, $T$ depends on the center-of-mass energy, system
size, centrality interval, jet type, jet order, channel of
detection system, and origination. Generally, $T$ increases with
increase of center-of-mass energy, that renders the larger the
center-of-mass energy is, the more violent transverse excitation
of the interacting system is. Besides, at the same center-of-mass
energy, there is a slightly negative correlation between $T$ and
the system size, or $T$ is approximately independent of the size
in the error range. The former case reflects that, in relatively
complex collision system, there is a jet quenching effect (quickly
energy loss) in the case of high energy quark and gluon jets
penetrating through the dense deconfined matter.

(d) From $p_T$ distributions of charged jets produced in Pb-Pb
collisions at $\sqrt{s_{NN}}=2.76$ TeV with different centrality
intervals analyzed by using the two-component Erlang distribution,
we know that $T$ slightly increases with increase of the
centrality percentage $C$ and quickly saturates in semi-central
collisions, or $T$ has no obvious change in the error range with
increase of $C$. In addition, as expected, the extracted effective
temperature from the reconstructed jets $p_T$ spectra produced in
electron channels in $p$-$p$ collisions at $\sqrt{s}=7$ TeV
decreases with increase of the jet order from the $1^{st}$ to
$5^{th}$.

(e) At the same energy, the effective temperature extracted from
the leading jets is much higher than that from the subleading
jets. This is a natural result because the leading jets are
produced through more violent scattering. The effective
temperature extracted from tagged jets (such as light-quark jets
and $b$-quark jets) seems to be higher than that from non-tagged
jets, which reveals that the hadron jets originated from quark
jets carry more energy and the leading jets primarily originate
from quark jets. The effective temperatures extracted from
different lepton and dilepton channel jets are approximately the
same in the error range, which reflects the common property in
these jets.
\\

{\bf Conflict of Interests}

The authors declare that there is no conflict of interests
regarding the publication of this paper.
\\

{\bf Acknowledgments}

One of the authors (Fu-Hu Liu) thanks Prof. Dr. Charles Gale,
Prof. Dr. Sangyong Jeon, and the members of the Physics Department
of McGill University, Canada, for their hospitality, where this
work was partly finished. The authors' work was supported by the
National Natural Science Foundation of China under Grant No.
10975095, the Open Research Subject of the Chinese Academy of
Sciences Large-Scale Scientific Facility under Grant No. 2060205,
the Shanxi Provincial Natural Science Foundation under Grant No.
2013021006, and the Foundation of Shanxi Scholarship Council of
China under Grant No. 2012-012.

\newpage

\renewcommand{\baselinestretch}{0.6}

{\small {Table 1. Values of free parameters, defined $T$, and
$\chi^2$/dof ($\chi^2$) corresponding to the curves in the
figures, where $\chi^2$ is given in bracket if the number of data
points is less than 6 which is the number of free parameters and
normalization constant. The errors of $m_j$ are estimated to be 0.
{%
\begin{center}
\begin{tabular}{cccccccc}
\hline\hline  Figure, type & $m_{1}$ & $p_{ti1}$ (GeV/$c$) & $k_{1}$ & $m_{2}$ & $p_{ti2}$ (GeV/$c)$ & $T$ (GeV) & $\chi^2$/dof ($\chi^2$)\\
\hline
1(a) & 2  & $1.70\pm0.20$  & $0.860\pm0.030$ & 2 & $3.25\pm0.10$    & $1.917\pm0.205$  & 0.328 \\
1(b) & 2  & $1.60\pm0.20$  & $0.900\pm0.030$ & 2 & $3.55\pm0.20$    & $1.795\pm0.215$  & 0.189\\
1(c) & 2  & $1.82\pm0.07$  & $0.994\pm0.002$ & 2 & $5.00\pm0.50$    & $1.839\pm0.070$  & 0.038 \\
1(d) & 11 & $1.25\pm0.05$  & $0.890\pm0.020$ & 8 & $2.47\pm0.05$    & $1.384\pm0.071$  & (0.084)\\
2(a) & 2  & $8.80\pm0.80$  & $0.830\pm0.030$ & 2 & $22.80\pm1.50$   & $11.180\pm1.022$ & 0.132\\
2(b) & 2  & $8.80\pm0.70$  & $0.840\pm0.030$ & 2 & $23.00\pm1.50$   & $11.072\pm0.974$ & 0.191\\
2(c) & 2  & $10.20\pm1.00$ & $0.900\pm0.030$ & 2 & $26.60\pm2.00$   & $11.840\pm1.257$ & 0.393\\
2(d) & 2  & $12.00\pm8.00$ & $0.996\pm0.001$ & 2 & $34.00\pm0.50$   & $12.088\pm7.968$ & 0.186\\
3(a) & 2  & $5.20\pm0.04$  & $0.900\pm0.020$ & 2 & $11.20\pm0.40$   & $5.800\pm0.253$  & 0.044\\
3(b) & 2  & $5.50\pm0.03$  & $0.950\pm0.010$ & 2 & $12.70\pm0.40$   & $5.860\pm0.143$  & 0.097\\
3(c) & 2  & $5.50\pm0.02$  & $0.950\pm0.010$ & 2 & $12.80\pm0.50$   & $5.865\pm0.143$  & 0.011\\
3(d) & 2  & $5.60\pm0.03$  & $0.960\pm0.010$ & 2 & $12.70\pm0.80$   & $5.884\pm0.145$  & 0.054\\
4(a), $1^{st}$ & 2  & $13.00\pm2.00$ & $0.620\pm0.080$ & 2 & $40.00\pm5.00$   & $23.260\pm4.058$ & 0.100 \\
4(a), $2^{nd}$ & 2  & $11.50\pm1.00$ & $0.850\pm0.050$ & 2 & $28.00\pm5.00$   & $13.975\pm1.891$ & 0.185 \\
4(b), lepton   & 2  & $28.00\pm2.00$ & $0.907\pm0.010$ & 2 & $100.00\pm30.00$ & $30.160\pm2.377$ & 0.019 \\
4(b), dilepton & 2  & $25.00\pm2.00$ & $0.850\pm0.050$ & 2 & $47.00\pm5.00$   & $28.300\pm3.246$ & 0.055 \\
4(c), inclusive& 2  & $42.00\pm2.00$ & $0.998\pm0.001$ & 2 & $115.00\pm10.00$ & $42.164\pm2.000$ & 0.619 \\
4(c), split    & 2  & $34.00\pm3.00$ & $0.994\pm0.003$ & 2 & $87.00\pm10.00$  & $34.318\pm2.996$ & 3.769 \\
4(d), lepton   & 2  & $21.90\pm2.00$ & $0.890\pm0.090$ & 2 & $40.50\pm10.00$  & $23.946\pm4.642$ & 0.095 \\
4(d), dilepton & 2  & $22.50\pm2.00$ & $0.880\pm0.100$ & 2 & $47.00\pm10.00$  & $25.440\pm5.629$ & 0.392 \\
5(a) & 2  & $13.50\pm1.00$ & $0.910\pm0.050$ & 2 & $50.00\pm20.00$  & $16.785\pm3.282$ & 0.303\\
5(b) & 2  & $11.20\pm1.00$ & $0.820\pm0.100$ & 2 & $28.00\pm10.00$  & $14.224\pm3.606$ & 0.312 \\
5(c) & 2  & $18.00\pm2.00$ & $0.810\pm0.100$ & 2 & $29.00\pm5.00$   & $20.090\pm3.896$ & 0.067 \\
6(a) & 4  & $12.70\pm0.90$ & $0.900\pm0.020$ & 2 & $44.00\pm4.00$   & $15.830\pm1.286$ & 0.187 \\
6(b) & 6  & $11.50\pm1.00$ & $0.710\pm0.100$ & 2 & $80.00\pm30.00$  & $31.365\pm11.896$& 0.661 \\
6(c) & 4  & $18.80\pm1.00$ & $0.960\pm0.010$ & 2 & $92.00\pm10.00$  & $21.728\pm1.401$ & 0.181\\
6(d) & 5  & $15.20\pm0.80$ & $0.770\pm0.080$ & 2 & $42.00\pm5.00$   & $21.364\pm3.804$ & 0.173\\
7, leading  & 7  & $14.00\pm1.00$ & $0.610\pm0.050$ & 2 & $58.00\pm2.00$   & $31.160\pm3.143$ & (0.069) \\
7, $2^{nd}$ & 7  & $9.60\pm0.08$  & $0.740\pm0.050$ & 2 & $42.00\pm2.00$   & $18.024\pm2.217$ & (0.367) \\
7, $3^{rd}$ & 7  & $6.10\pm0.50$  & $0.620\pm0.050$ & 2 & $23.00\pm1.00$   & $12.522\pm1.287$ & (0.434) \\
7, $4^{th}$ & 7  & $4.70\pm1.00$  & $0.780\pm0.050$ & 2 & $18.80\pm1.50$   & $7.802\pm1.287$  & (0.281) \\
7, $5^{th}$ & 7  & $4.00\pm0.70$  & $0.820\pm0.040$ & 2 & $15.00\pm1.50$   & $5.980\pm0.888$  & (0.128) \\
8(a) & 5  & $22.00\pm2.00$ & $0.820\pm0.030$ & 2 & $82.00\pm4.00$   & $32.800\pm3.114$ & 0.177 \\
8(b) & 6  & $17.60\pm1.00$ & $0.880\pm0.030$ & 2 & $79.00\pm5.00$   & $24.968\pm2.651$ & 0.619 \\
8(c) & 5  & $23.00\pm3.00$ & $0.580\pm0.050$ & 2 & $81.00\pm5.00$   & $47.360\pm5.016$ & 0.408 \\
8(d) & 6  & $18.00\pm1.00$ & $0.820\pm0.040$ & 2 & $100.00\pm10.00$ & $32.760\pm4.520$ & 0.799 \\
9(a) & 4  & $24.90\pm2.00$ & $0.880\pm0.040$ & 2 & $89.00\pm7.00$   & $32.592\pm4.180$ & 1.022 \\
9(b) & 6  & $36.50\pm1.00$ & $0.850\pm0.030$ & 2 & $132.00\pm8.00$  & $50.825\pm4.364$ & 0.112 \\
9(c) & 6  & $18.50\pm1.00$ & $0.880\pm0.050$ & 2 & $120.00\pm50.00$ & $30.680\pm8.581$ & 0.427 \\
9(d) & 6  & $19.10\pm1.00$ & $0.930\pm0.030$ & 2 & $160.00\pm50.00$ & $28.963\pm6.040$ & 0.453 \\
9(e) & 7  & $16.00\pm1.00$ & $0.870\pm0.050$ & 2 & $109.00\pm15.00$ & $28.090\pm5.908$ & 0.815 \\
9(f) & 9  & $11.60\pm1.00$ & $0.750\pm0.050$ & 2 & $180.00\pm100.00$& $53.700\pm26.588$& 0.731 \\
10(a) & 9 & $11.00\pm0.70$ & $0.700\pm0.070$ & 4 & $41.00\pm3.00$   & $20.000\pm3.143$ & 0.196 \\
10(b) & 9 & $10.90\pm0.70$ & $0.710\pm0.050$ & 4 & $40.00\pm2.00$   & $19.339\pm2.209$ & 0.138\\
10(c) & 5 & $15.20\pm2.00$ & $0.540\pm0.100$ & 2 & $69.00\pm10.00$  & $39.948\pm8.500$ & 0.137 \\
10(d) & 5 & $5.80\pm1.00$  & $0.580\pm0.100$ & 2 & $26.00\pm5.00$   & $14.284\pm3.441$ & 0.298 \\
10(e) & 2 & $25.30\pm2.00$ & $0.830\pm0.030$ & 2 & $61.00\pm3.00$   & $31.369\pm2.634$ & 0.053 \\
10(f) & 2 & $12.50\pm0.50$ & $0.970\pm0.003$ & 2 & $47.00\pm2.00$   & $12.877\pm0.507$ & 0.295\\
11(a) & 4  & $45.80\pm4.00$ & $0.969\pm0.010$ & 2 & $118.00\pm5.00$  & $48.038\pm4.080$ & 0.392 \\
11(b) & 6  & $3.00\pm0.10$  & $0.760\pm0.020$ & 2 & $19.00\pm1.00$   & $6.840\pm0.460$  & 0.062 \\

\hline\hline
\end{tabular}%
\end{center}
}} }

\renewcommand{\baselinestretch}{1.}

\newpage
\begin{figure}
\hskip-1.0cm \begin{center}
\includegraphics[width=12.0cm]{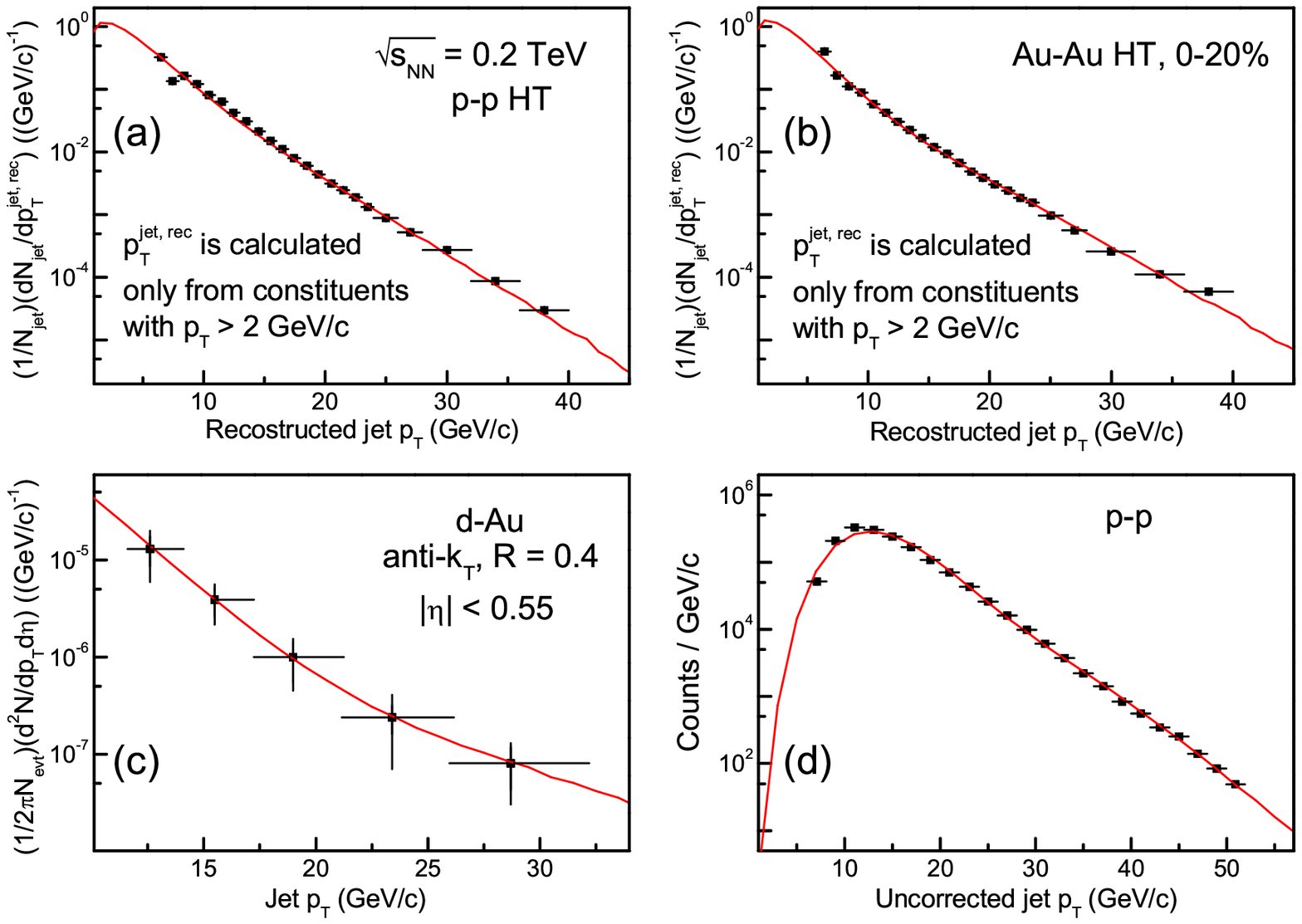}
\end{center}
\vskip-1.0cm Fig. 1. Transverse momentum spectra of HT trigger
jets in $p$-$p$ collision (a), HT trigger jets in 0--20\% central
Au-Au collisions (b), jets from $d$-Au collisions (c), and
uncorrected jets in $p$-$p$ collision (d) at $\sqrt{s_{NN}}=0.2$
TeV. The symbols represent the experimental data of the STAR
Collaboration [9--11] and the curves are our results calculated by
using the two-component Erlang distribution in the framework of
the multi-source thermal model. All the calculations are performed
by the Monte Carlo method.
\end{figure}

\begin{figure}
\hskip-1.0cm \begin{center}
\includegraphics[width=12.0cm]{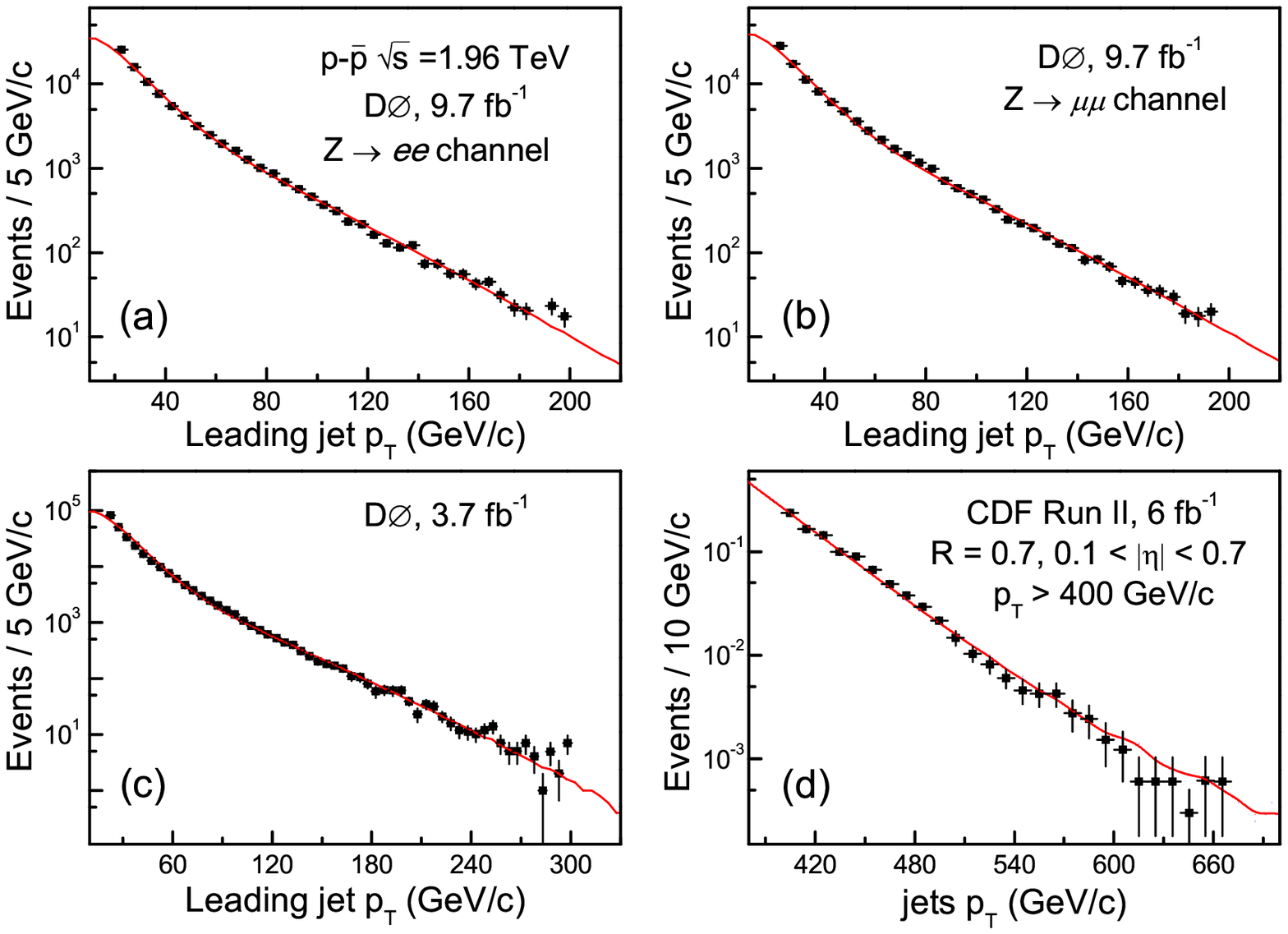}
\end{center}
\vskip-1.0cm Fig. 2. Transverse momentum spectra for the leading
jets (a)--(c) and all the jet candidates (d) produced in
$p$-$\bar{p}$ collision at $\sqrt{s}=1.96$ TeV. The symbols in
Figs. 2(a)/2(b) and Figs. 2(c)/2(d) represent the experimental
data of the D0 Collaboration [13, 14] and the CDF II Collaboration
[15] respectively. The curves are our results calculated by using
the two-component Erlang distribution.
\end{figure}

\newpage
\begin{figure}
\hskip-1.0cm \begin{center}
\includegraphics[width=12.0cm]{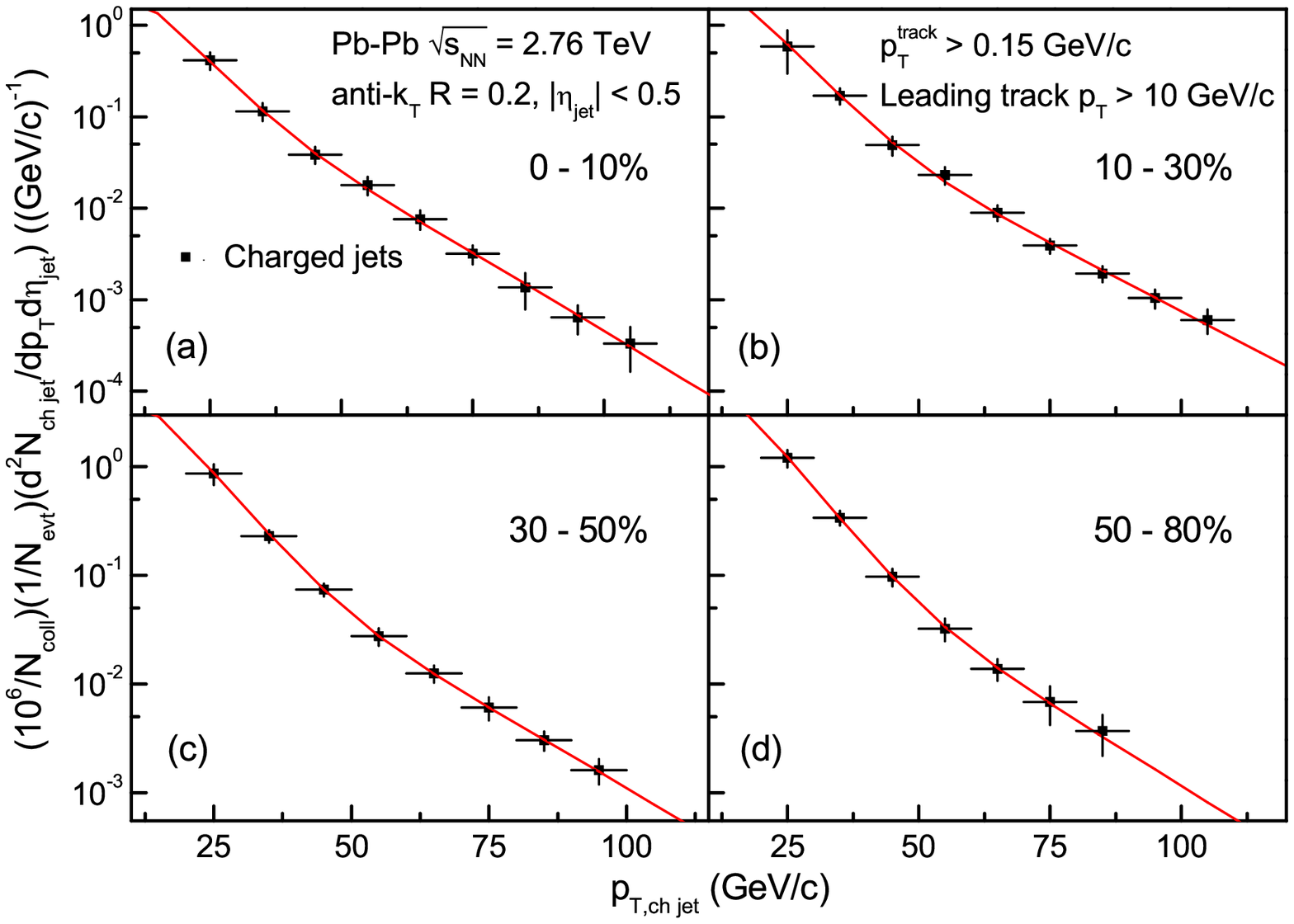}
\end{center}
\vskip-.50cm Fig. 3. Charged jet $p_T$ spectra produced in Pb-Pb
collisions at $\sqrt{s_{NN}}=2.76$ TeV in different centrality
intervals of 0--10\% (a), 10--30\% (b), 30--50\% (c), and 50--80\%
(d). The symbols represent the experimental data of the ALICE
Collaboration [16] and the curves are our results calculated by
using the two-component Erlang distribution.
\end{figure}

\newpage
\begin{figure}
\hskip-1.0cm \begin{center}
\includegraphics[width=13.0cm]{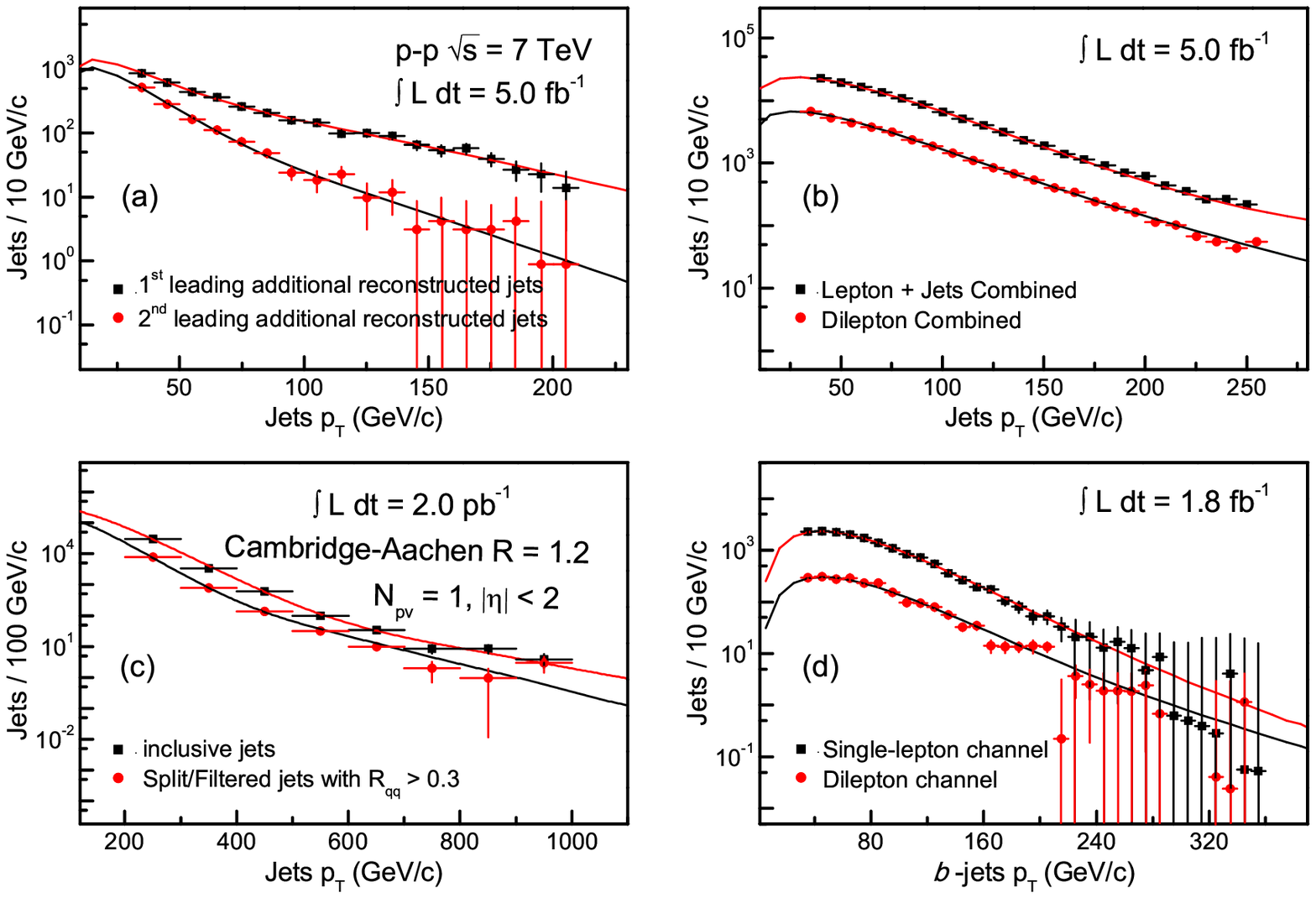}
\end{center}
\vskip-1.0cm Fig. 4. Jet $p_T$ spectra produced in $p$-$p$
collision at $\sqrt{s}=7$ TeV for different situations shown in
the panels and text. The symbols shown in Figs. 4(a)/4(b) and
Figs. 4(c)/4(d) represent the experimental data of the CMS
Collaboration [17] and the ATLAS Collaboration [18, 19]
respectively. The curves are our modelling results.
\end{figure}

\newpage
\begin{figure}
\hskip1.0cm \begin{center}
\includegraphics[width=16.0cm]{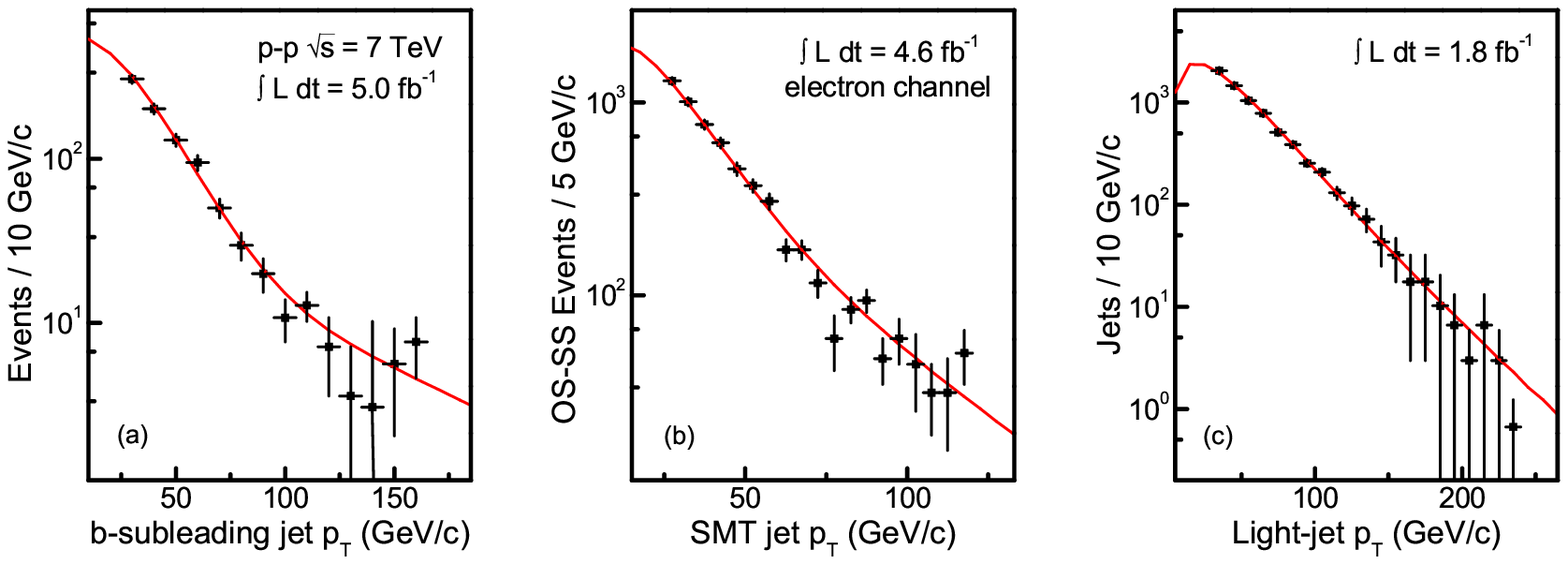}
\end{center}
\vskip-0.3cm Fig. 5. Transverse momentum spectra for the
subleading $b$-jets (a), the SMT jets (b), and the light-quark
jets (c) produced in $p$-$p$ collision at $\sqrt{s}=7$ TeV. The
symbols represent the experimental data of the CMS Collaboration
[20] and the ATLAS Collaboration [19, 21]. The curves are our
modelling results.
\end{figure}

\newpage
\begin{figure}
\hskip-1.0cm \begin{center}
\includegraphics[width=15.0cm]{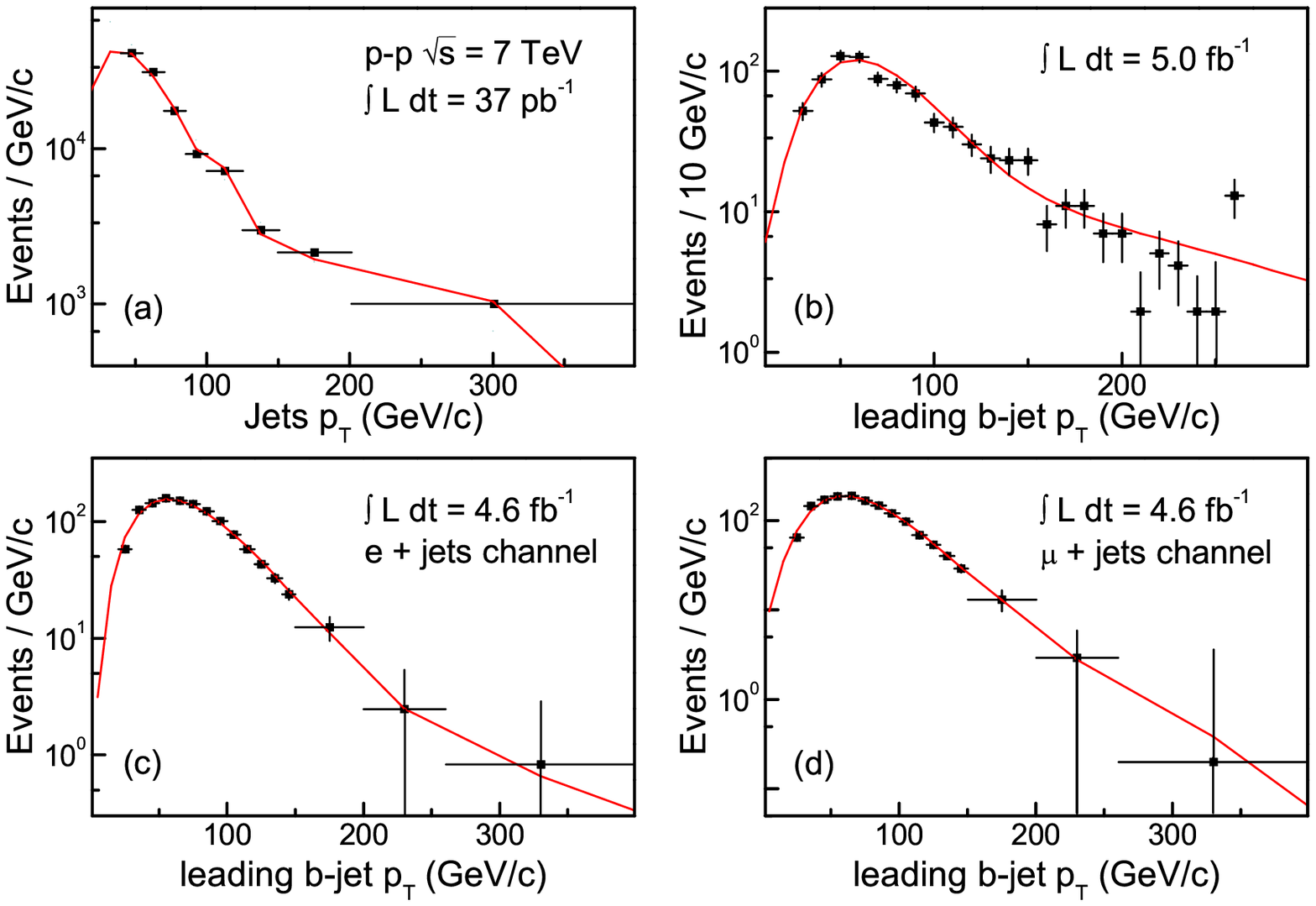}
\end{center}
\vskip-1.0cm Fig. 6. Some other jet $p_T$ spectra in $p$-$p$
collision at $\sqrt{s}=7$ TeV for different situations shown in
the panels and text. The symbols shown in Figs. 6(a), 6(b), and
6(c)/6(d) represent the experimental data of the ATLAS [22], CMS
[20], and ATLAS Collaborations [23], respectively. The curves are
our modelling results.
\end{figure}

\newpage
\begin{figure}
\hskip-1.0cm \begin{center}
\includegraphics[width=10.0cm]{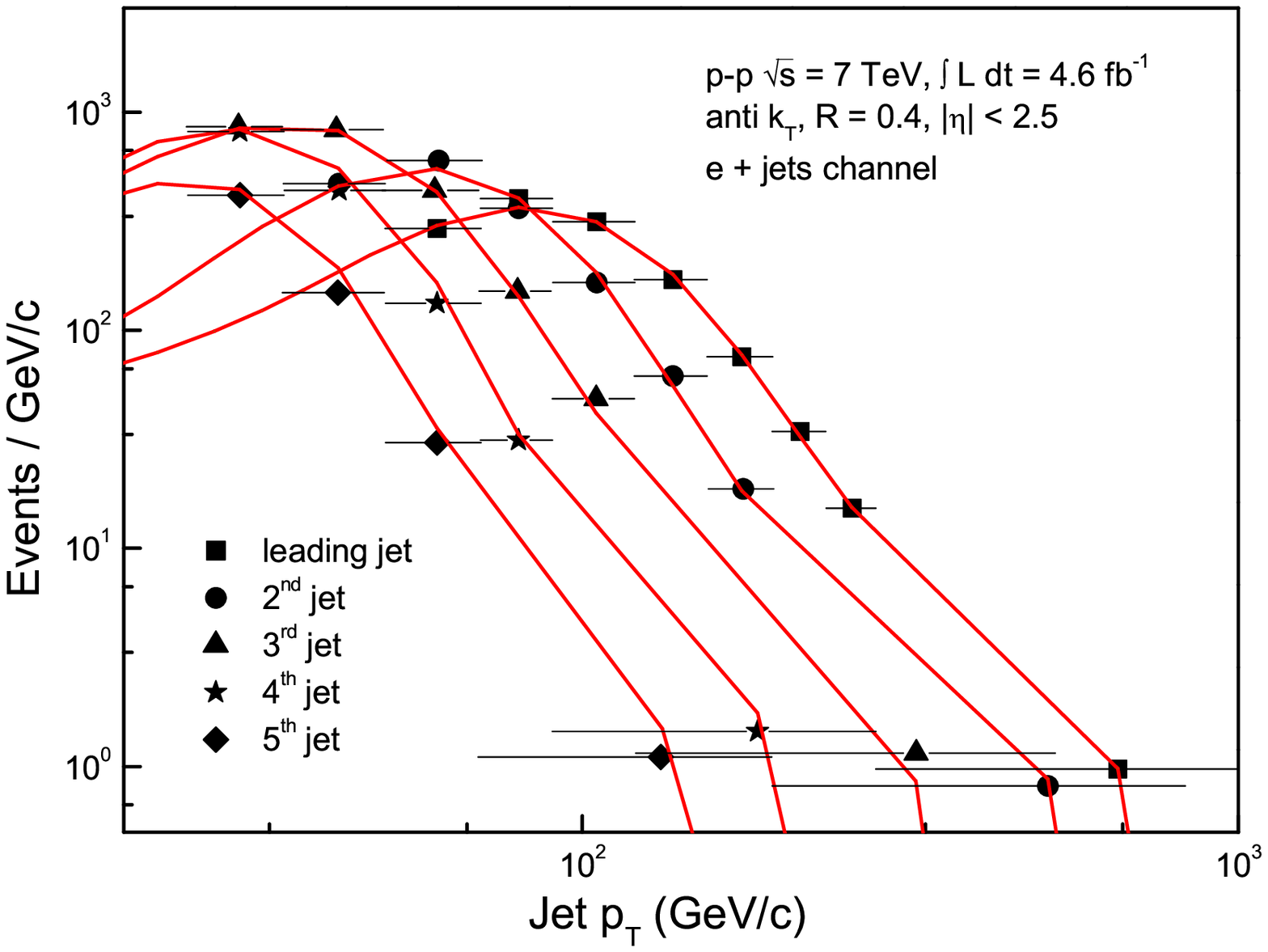}
\end{center}
\vskip-.60cm Fig. 7. The transverse momentum spectra for the
leading, $2^{nd}$ , $3^{rd}$, $4^{th}$, and $5^{th}$ order jets in
the electron ($e$+jets) channels produced in $p$-$p$ collision at
$\sqrt{s}=7$ TeV. The symbols represent the experimental data of
the ATLAS Collaboration [24] and the curves are our modelling
results.
\end{figure}

\newpage
\begin{figure}
\hskip-1.0cm \begin{center}
\includegraphics[width=14.0cm]{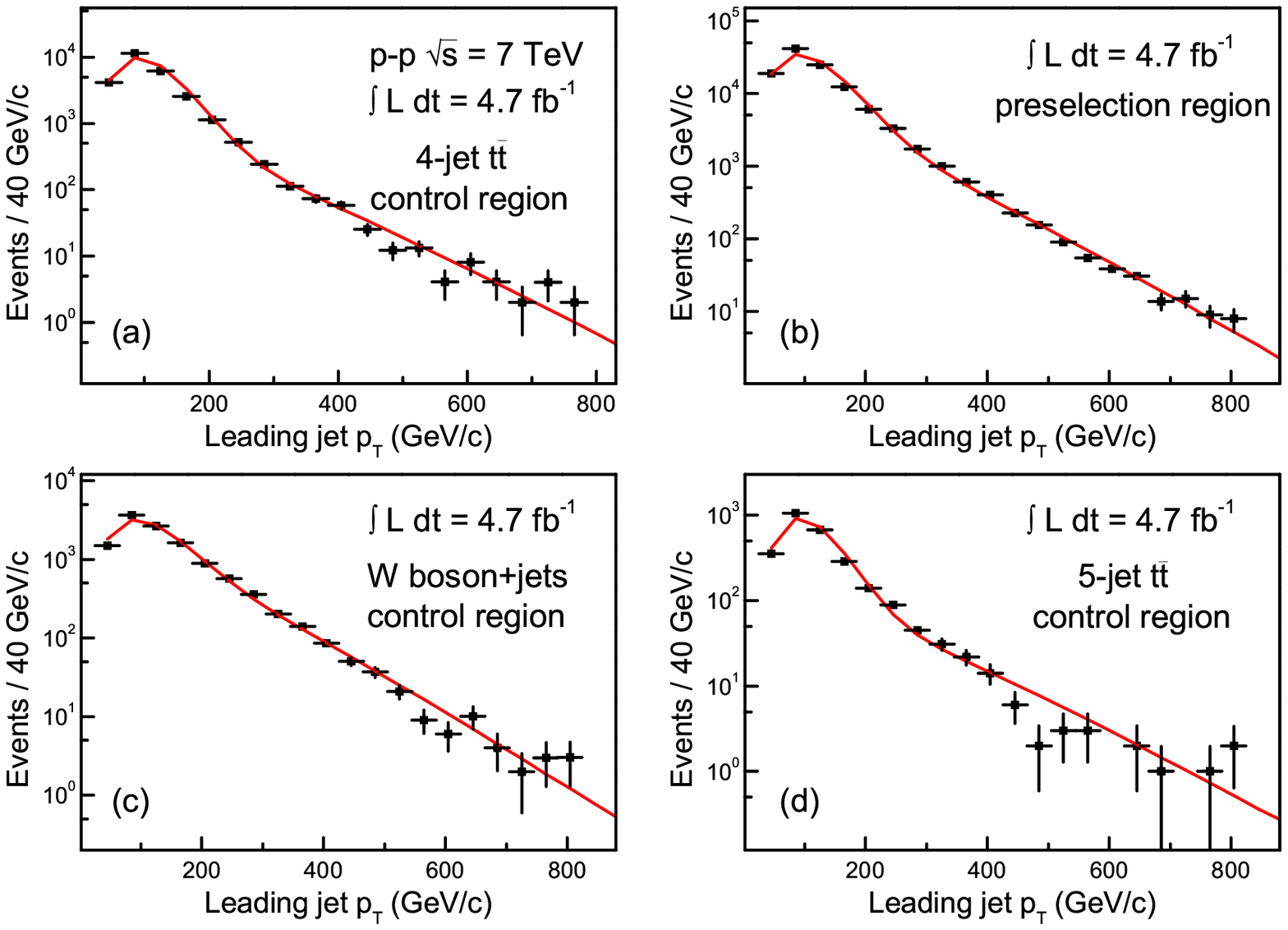}
\end{center}
\vskip-1.0cm Fig. 8. The leading jet $p_T$ distributions in (a)
the 4-jet $t \bar{t}$, (b) preselection, (c) $W$ boson+jets, and
(d) 5-jet $t\bar{t}$ with an integrated luminosity of 4.7
fb$^{-1}$ in $p$-$p$ collision at $\sqrt{s}=7$ TeV. The symbols
represent the experimental data of the ATLAS Collaboration [25]
and the curves are our modelling results.
\end{figure}

\newpage
\begin{figure}
\hskip-1.0cm \begin{center}
\includegraphics[width=12.0cm]{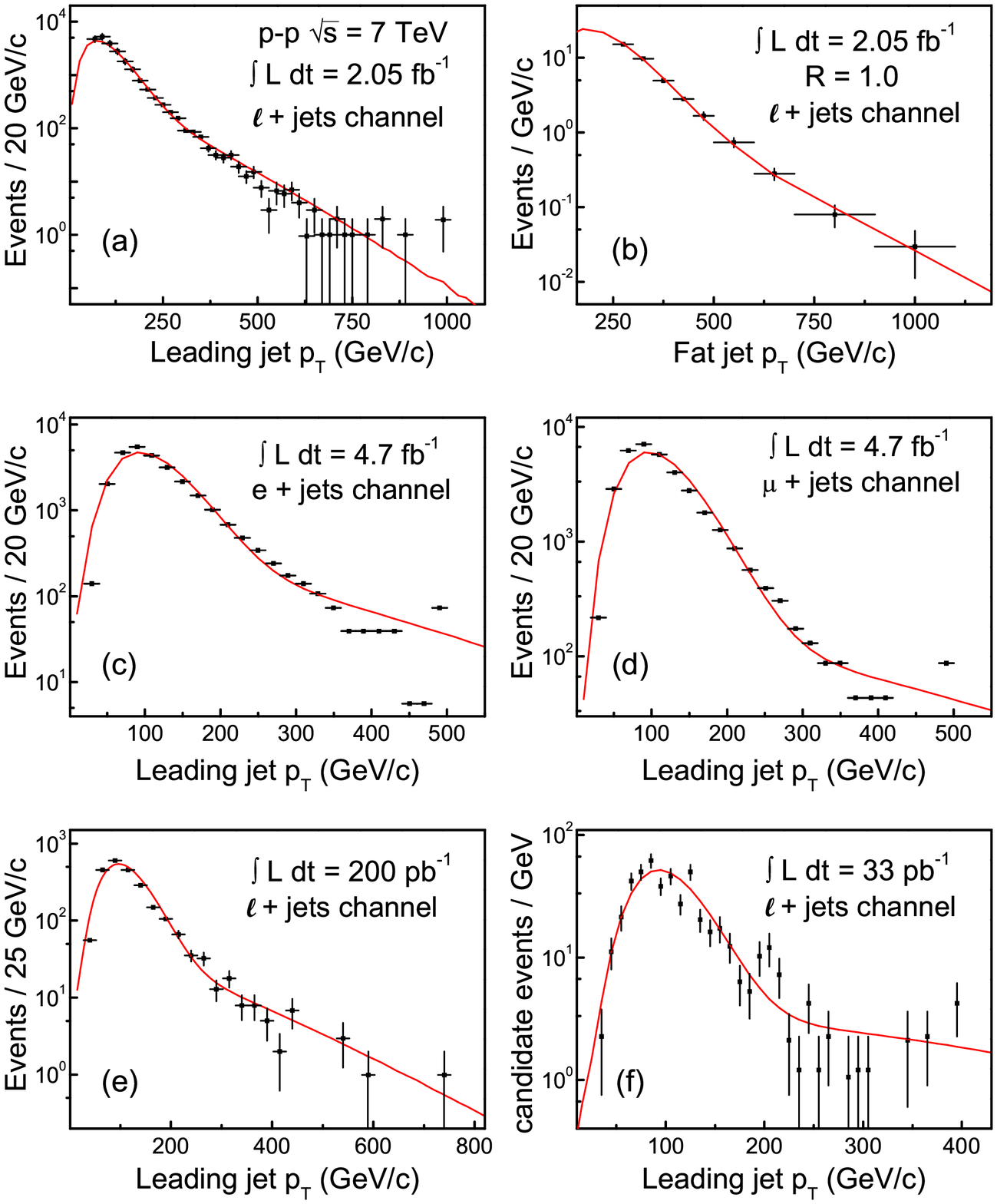}
\end{center}
\vskip-1.0cm Fig. 9. Transverse momentum distributions of the
leading and fat jets produced in $p$-$p$ collision at $\sqrt{s}=7$
TeV with different channels and integrated luminosities shown in
the panels and text. The symbols represent the experimental data
of the ATLAS Collaboration [26--30] and the curves are our
modelling results.
\end{figure}

\newpage
\begin{figure}
\hskip-1.0cm \begin{center}
\includegraphics[width=13.0cm]{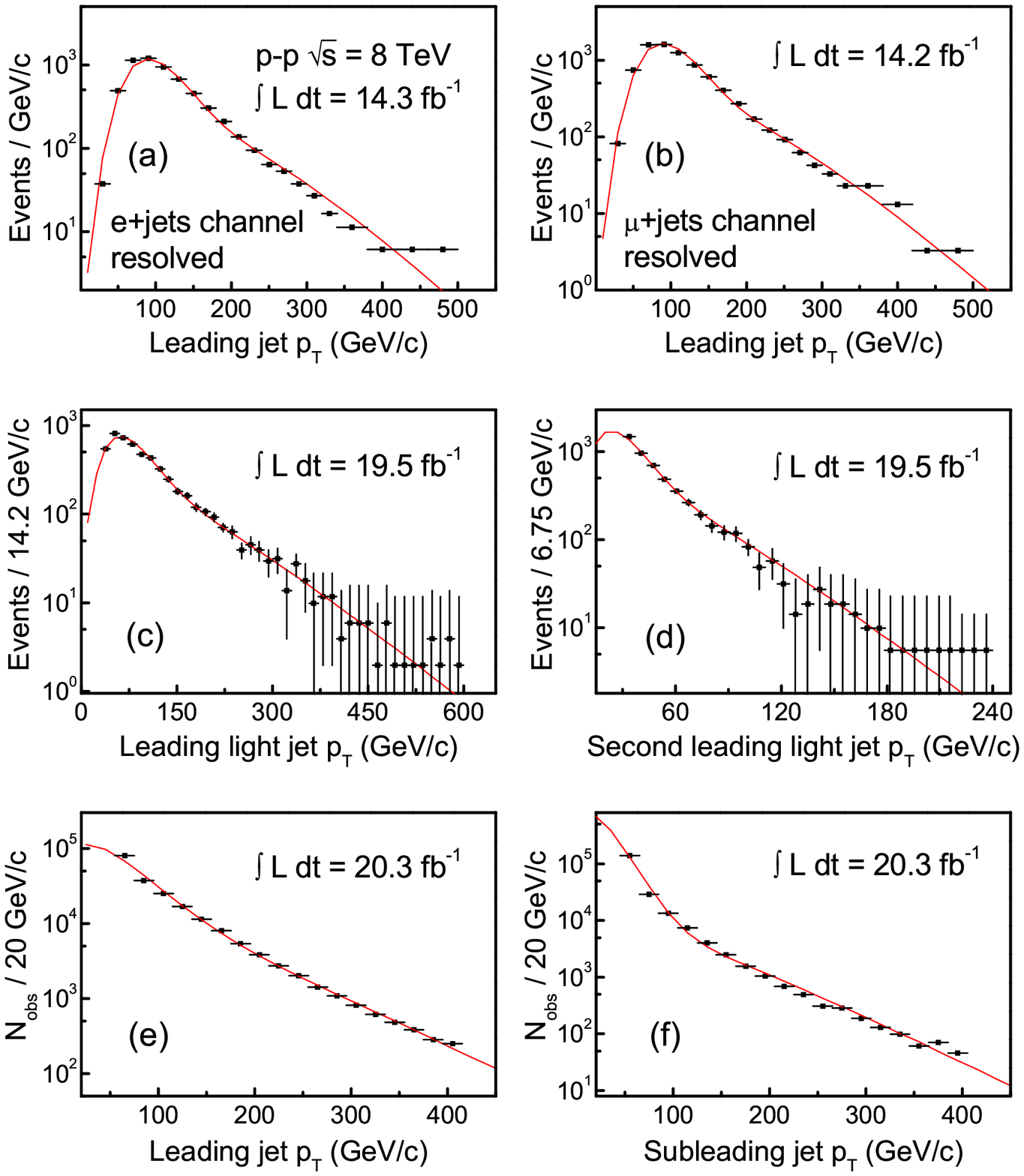}
\end{center}
\vskip-1.0cm Fig. 10. Transverse momentum distributions of
different types of jets produced in $p$-$p$ collision at
$\sqrt{s}=8$ TeV with different selected conditions shown in the
panels and text. The symbols shown in Figs. 10(a)/10(b),
10(c)/10(d), and 10(e)/10(f) represent the experimental data of
the ATLAS [31], CMS [32], and ATLAS Collaborations [33],
respectively. The curves are our modelling results.
\end{figure}

\begin{figure}
\hskip-1.0cm \begin{center}
\includegraphics[width=16.0cm]{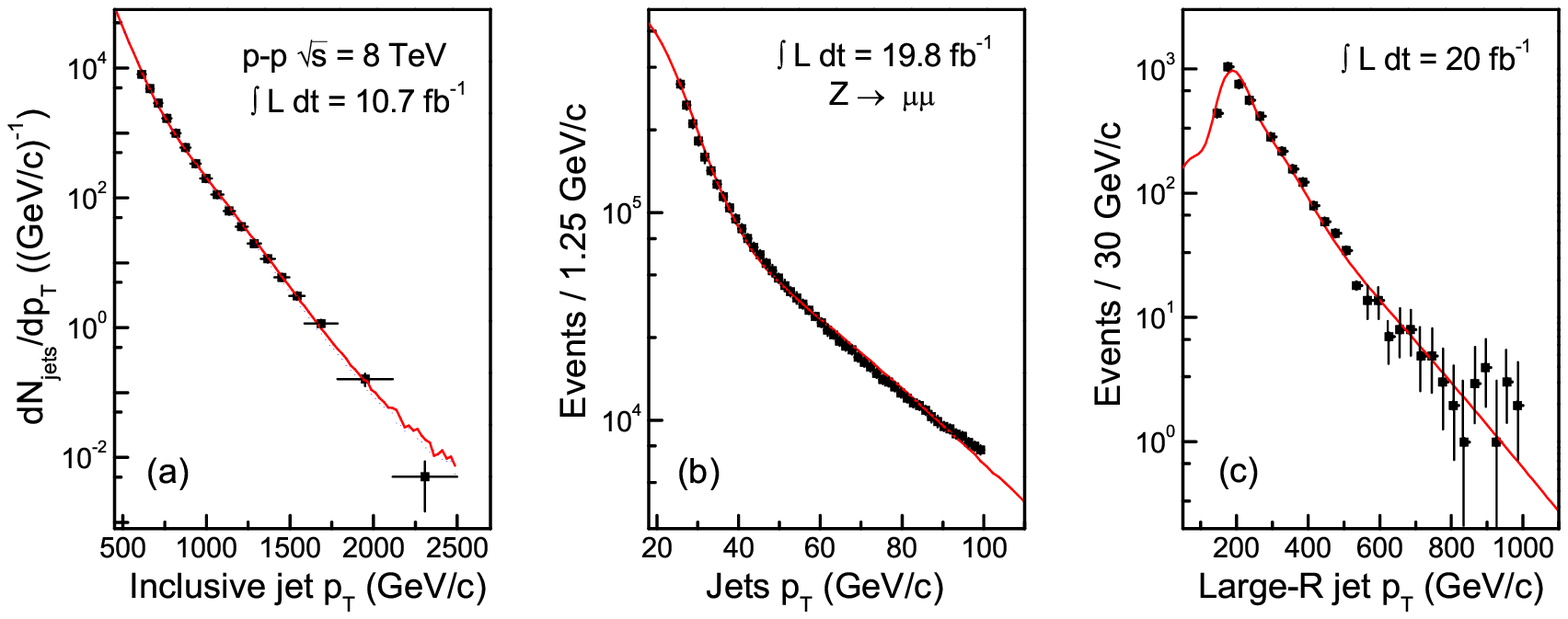}
\end{center}
\vskip-.5cm Fig. 11. (a) and (b) Inclusive jet $p_T$ spectra and
(c) the large-$R$ jet $p_T$ spectrum produced in $p$-$p$ collision
at $\sqrt{s}=8$ TeV. The symbols shown in Figs. 11(a)/11(b) and
11(c) represent the experimental data of the CMS [34, 35] and
ATLAS Collaborations [36], respectively. The curves are our
modelling results, where the two-component Erlang distribution is
used for Figs. 11(a) and 11(b), and the three-component Erlang
distribution is used for Fig. 11(c).
\end{figure}

\newpage
\begin{figure}
\hskip-1.0cm \begin{center}
\includegraphics[width=8.0cm]{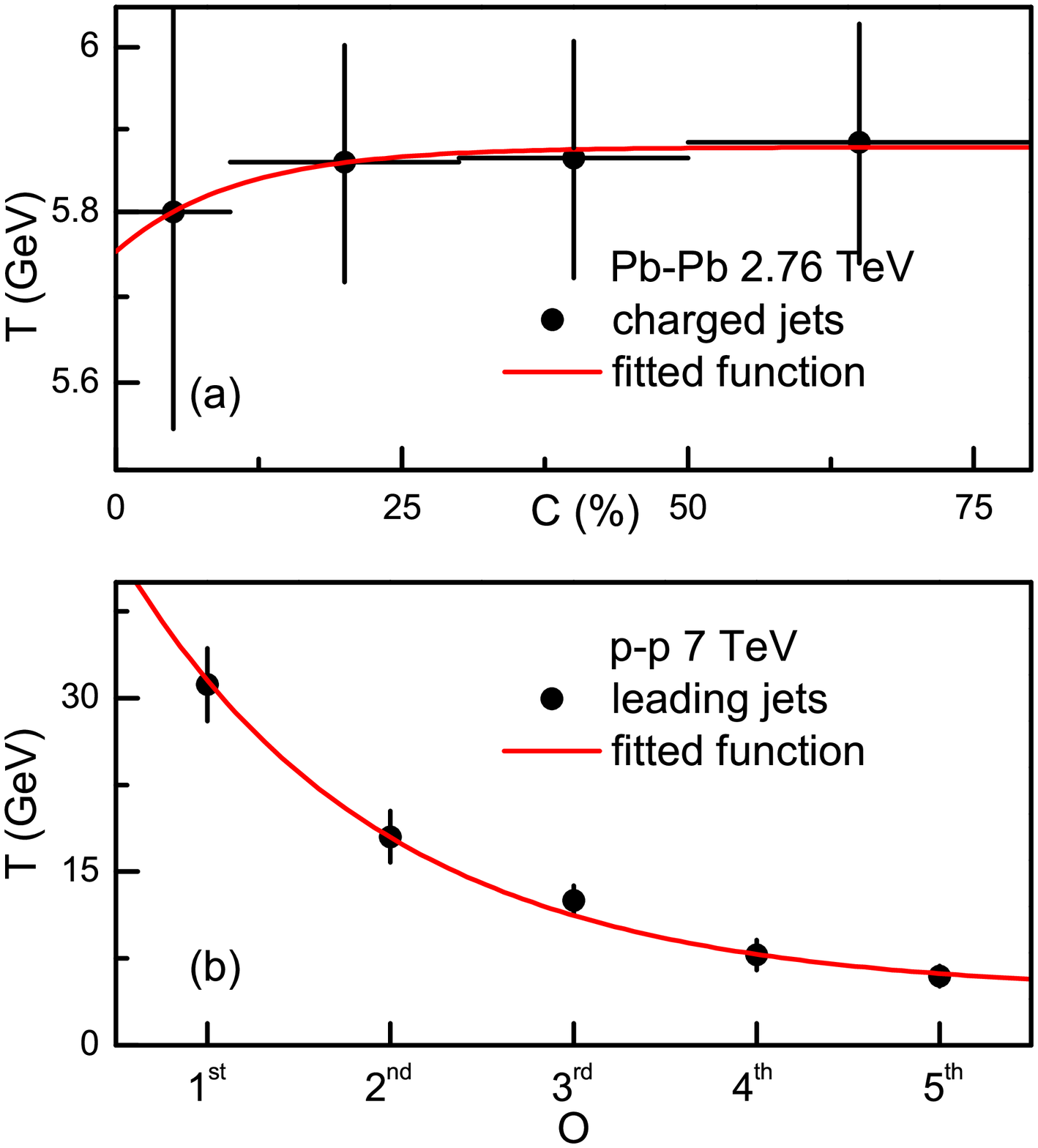}
\end{center}
\vskip-1.0cm Fig. 12. Dependences of defined effective temperature
parameter $T$ on centrality percentage $C$ in Pb-Pb collisions at
$\sqrt{s_{NN}}=2.76$ TeV (a) and on jet order $O$ in $p$-$p$
collision at $\sqrt{s}=7$ TeV (b). The symbols represent $T$
values obtained from Figs. 3 and 7 (listed in Table 1)
respectively, and the curves are our fitted results which are
presented in Eqs. (7) and (8) respectively.
\end{figure}

\newpage
\begin{figure}
\hskip-1.0cm \begin{center}
\includegraphics[width=15.0cm]{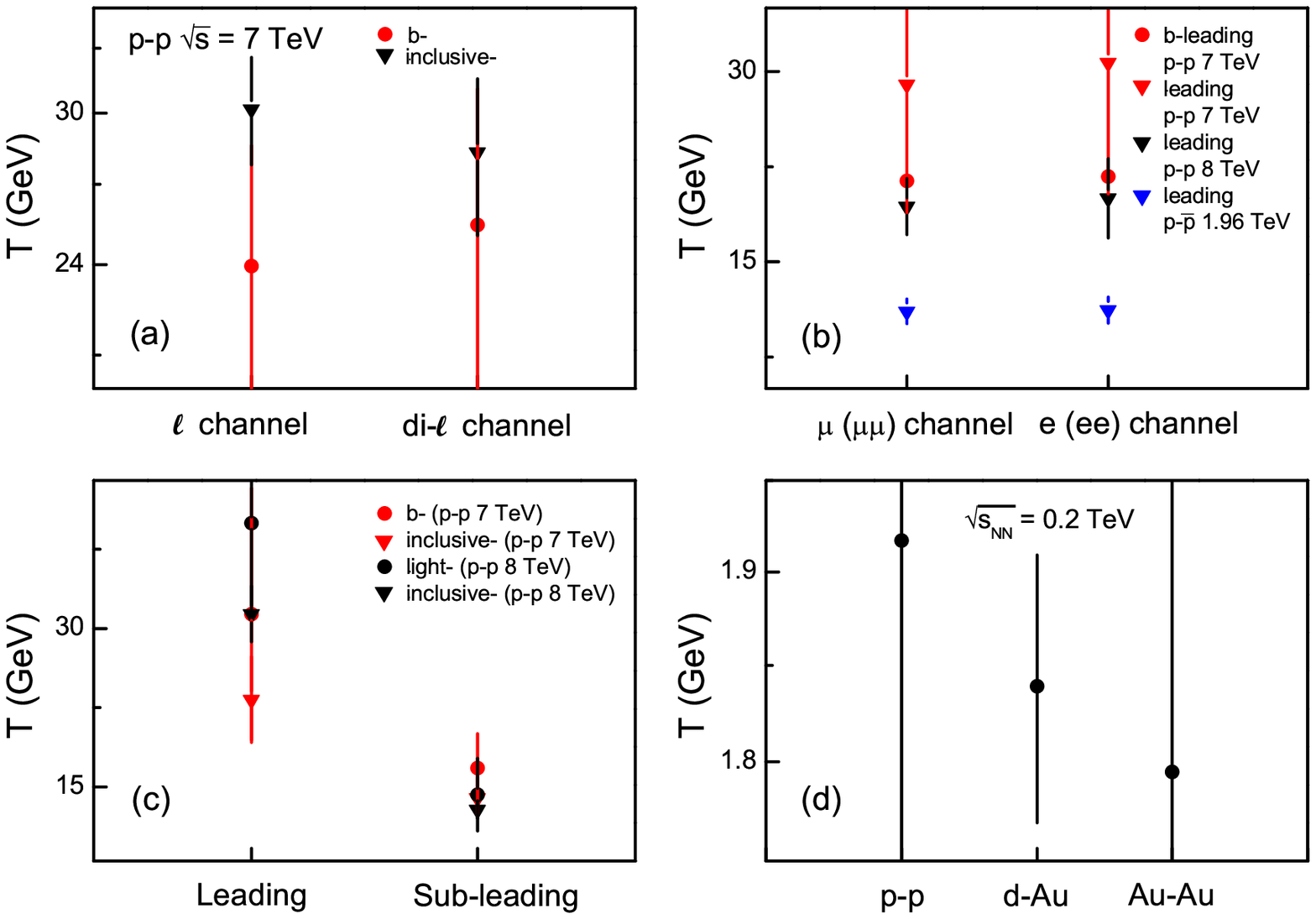}
\end{center}
\vskip-1.0cm Fig. 13. Dependences of $T$ on $l$ and di-$l$
channels (a), $\mu(\mu\mu)$ and $e(ee)$ channels (b), leading and
sub-leading jets (c), as well as size of interacting system (d).
The symbols represent $T$ values obtained from related figures and
listed in Table 1.
\end{figure}

\end{document}